\documentclass[twoside]{article}
\usepackage{fleqn,espcrc2,epsf}

\title{High Energy Cosmic Tau Neutrinos\thanks{Talk given at 
       6th International Workshop on Topics in Astroparticle and 
       Underground Physics (TAUP 99), 
        6-10 September, 1999, Paris, France.}}

\author{Athar Husain\\
\vspace{1pc}        
        Department of Physics, Tokyo Metropolitan University,\\
        Minami-Osawa 1-1, Hachioji-Shi, Tokyo 192-0397, Japan}

\begin{document}

\begin{abstract}

	I discuss the possibility of production of high energy cosmic tau 
 neutrinos ($E\, \geq 10^{6}$ GeV) in an astrophysical site and study some of 
 the effects of neutrino mixing on their subsequent propagation. 
 I also discuss the prospects for observations of these high energy cosmic 
 tau neutrinos through double shower events in new km$^{2}$ surface area 
 under water/ice neutrino telescopes.

\vspace{1pc}
\end{abstract}

\maketitle

\section{INTRODUCTION}

	In this contribution  i discuss the possibility of production of 
high energy cosmic tau neutrinos ($E\, \geq 10^{6}$ GeV)
 in cores of Active Galactic
Nuclei (AGN) originating from proton acceleration. The effects of three 
 flavour neutrino mixing on this high 
energy cosmic tau neutrino flux and the prospects for its observations 
through double shower events \cite{sandip}.

	In addition to AGNs, high energy cosmic tau neutrinos may also be 
produced in several currently envisaged other cosmologically distant
astrophysical sources \cite{taup}. 
 For some possible effects of neutrino 
mixing other than the flavour one on high energy cosmic tau neutrino flux, 
 see \cite{npb}. For prospects of observations of high 
energy cosmic tau neutrinos other than the double shower technique, see 
\cite{upward}. The present study is particularly useful as several 
 under water/ice high energy
neutrino telescopes  are now at their rather advanced stages of 
development and deployment \cite{moscoso}. 

	I start in Section 2 with a brief description of possibility of 
production of high energy cosmic tau neutrinos relative to electron and muon
neutrinos and discuss in some detail the effect of three flavour neutrino
mixing on their subsequent propagation and further discuss the prospects for
their detection through double shower events. In Section 3, i summarize the 
results.

\section{HIGH ENERGY COSMIC TAU NEUTRINOS}

\subsection{Production}

	High energy cosmic tau neutrinos may either mainly be produced in 
$p\gamma $ or in $pp $ collisions in a cosmologically distant environment. 

In $p\gamma $ collisions,
high energy $\nu_{e}$ and $\nu_{\mu}$ are mainly produced through the 
resonant reaction $p+\gamma \rightarrow \Delta^{+}\rightarrow n+\pi^{+}$. 
The same collisions will give rise to a greatly suppressed 
high energy $\nu_{\tau}$ (and $\bar{\nu}_{\tau}$) flux ($\nu_{\tau}/
 \nu_{e, \mu}\, <\, 10^{-5}$)
mainly through the 
reaction $p+\gamma \, \rightarrow \, D^{+}_{S}+\Lambda^{0}+\bar{D}^{0}$. 
In $pp$ collisions, the $\nu_{\tau}$ flux may be obtained through 
$p+p\rightarrow D^{+}_{S}+X$.
The relatively small cross-section for $D^{+}_{S}$ production 
together with the low branching 
ratio into $\nu_{\tau}$ implies that the 
$\nu_{\tau}$ flux in $pp$ collisions is also suppressed up to 5 
orders of magnitude relative to $\nu_{e}$ and/or $\nu_{\mu}$ fluxes.

Thus, both in $p\gamma $ and in $pp$ collisions, the intrinsically produced 
tau neutrino flux is expected to be rather quite small, typically a factor 
less than $10^{-5}$ relative to electron and muon neutrino fluxes 
\cite{athar}. 

\subsection{Propagation}

Matter effects on vacuum neutrino flavour oscillations are
relevant if $G_{F}\rho /m_{N}\, \sim \, \delta m^{2}/2E$. Using $\rho $ 
given in Ref. \cite{szabo}, it follows that matter effects are absent for 
$\delta m^{2}\, \geq \, {\cal O} (10^{-10})$ eV$^{2}$. 
Matter effects are not expected to be important in the neutrino 
production regions around AGN and will not be further discussed here.

	In the framework of three flavour analysis, the flavour precession
probability from $\alpha $ to $\beta $ neutrino flavour is \cite{book} 

\begin{eqnarray}
 P_{\alpha \beta} & = & \sum^{3}_{i=1}|U_{\alpha i}|^{2}|U_{\beta i}|^{2}
                         \nonumber  \\
                  &   & +\sum_{i \neq j} U_{\alpha i}U^{\ast}_{\beta i}
                        U^{\ast}_{\alpha j}U_{\beta j}
                        \cos\left(\frac{2\pi L}{l_{ij}}\right),
\end{eqnarray}
where $\alpha, \beta = e, \mu, $ or $\tau $. $U$ is the 3$\times $3 MNS mixing 
matrix and will be used in usual notation with the standard 
 parameterization for  vanishing $\theta_{13}$ and  CP violating 
 phase \cite{pdg}. In Eq. (1), 
 $l_{ij}\simeq 4\pi E/\delta m^{2}_{ij}$ with $\delta m^{2}_{ij} = 
 |m^{2}_{i}-m^{2}_{j}|$ and 
$L$ is the distance between the source and the detector.

	Currently, the atmospheric muon and solar electron neutrino deficits 
 can be explained with 
oscillations among three active neutrinos \cite{nakahata}. 
 For this, typically, $\delta m^{2}
\sim {\cal O}(10^{-3})$ eV$^{2}$ and $\sin^{2}2\theta \sim {\cal O}(1)$ for 
the explanation of atmospheric muon neutrino deficit, 
whereas for the explanation of solar electron neutrino deficit, we may have  
$\delta m^{2} \sim {\cal O}(10^{-10})$ eV$^{2}$ and 
$\sin^{2}2\theta \sim {\cal O}(1)$ [just so solution] or $\delta m^{2}
\sim {\cal O}(10^{-5})$ eV$^{2}$ and $\sin^{2}2\theta \sim {\cal O}(10^{-2})$
[SMA (MSW)] or $\delta m^{2}
\sim {\cal O}(10^{-5})$ eV$^{2}$ and $\sin^{2}2\theta \sim {\cal O}(1)$ 
[LMA (MSW)]. 

	In the above explanations, the total range of $\delta m^{2}$ is 
$10^{-10}\leq \delta m^{2}/$ eV$^{2} \leq 10^{-3}$ irrespective of neutrino 
 flavour. The typical energy span relevant for possible flavour 
 identification for high
energy cosmic neutrinos is $2\cdot 10^{6}\leq E/$GeV$\leq 2\cdot 10^{7}$ in 
which currently the neutrino flux from cores of AGNs dominate. Taking a 
typical distance between the AGN and our galaxy as $L \sim 100$ Mpc (where 
1 pc $\sim 3\cdot 10^{16}$ m), the $\cos $ term in Eq. (1) vanishes 
and so Eq. (1) reduces to 
 
\begin{equation}
   \langle P_{\alpha \beta} \rangle \simeq  
   \sum^{3}_{i=1}|U_{\alpha i}|^{2}|U_{\beta i}|^{2}.
\end{equation}

It is assumed here that no relatively dense objects exist between the AGN and 
the earth so as to effect significantly this oscillations pattern. The 
$\langle P_{\alpha \beta } \rangle $ in Eq. (2) is independent of not only
$\delta m^{2}$ but also $E$. 

	Let me denote by $F^{0}_{\alpha }$, the intrinsic high energy cosmic 
 neutirno fluxes. For simplicity, i take their ratios as  
$F^{0}_{e} : F^{0}_{\mu} :F^{0}_{\tau} = 1: 2: 0$. In order to estimate the 
 final flux ratios 
on earth, let me  introduce a 3$\times $3 matrix of vacuum flavour 
precession probabilities such that 

\begin{equation}
 F_{\alpha} = \sum_{\beta}\langle P_{\alpha \beta}\rangle F^{0}_{\beta}.
\end{equation}

The explicit form for the matrix $\langle P \rangle $ in case of 
just so flavour oscillations as solution to solar neutrino problem along with
the solution to atmospheric neutrino deficit in terms of $\nu_{\mu}$ to 
$\nu_{\tau}$ oscillations with maximal depth is

\begin{equation}
 \langle P \rangle = \left( \begin{array}{ccc}
                             1/2 & 1/4 & 1/4 \\
                             1/4 & 3/8 & 3/8 \\
                             1/4 & 3/8 & 3/8 
                            \end{array}
                      \right).
\end{equation}
Using Eq. (4) and Eq. (3), we note that $F_{e}: F_{\mu }: F_{\tau } = 1:
1: 1$ at the level of $F^{0}_{e}$. Also, the unitarity conditions for 
 $\langle P_{\alpha \beta }\rangle $  are  satisfied. The same  
flux ratio is obtained in remaining two cases. Thus, essentially independent 
of the 
oscillation solution for solar neutrino problem, i  have 
 $F_{e}: F_{\mu }: F_{\tau } = 1: 1: 1$. This provides some prospects for 
possible detection of high energy cosmic tau neutrinos.

\subsection{Prospects for detection}

The down ward going high energy cosmic tau neutrinos reaching close to the 
surface of the detector may undergo  a charged current 
 deep inelastic 
scattering with nuclei inside/near the detector and produce 
 a tau lepton in addition
to a hadronic shower. This tau lepton
traverses a distance, on average proportional to its energy, 
before it decays back into a tau 
neutrino and a second shower most often induced by decay hadrons. 
The second shower is expected to carry about twice as much energy 
as the first and such double shower signals are commonly referred 
to as a double bangs. 
As tau leptons are not expected to have further relevant 
interactions (with high energy loss) in their decay 
time scale, the two showers should be 
 separated by a clean $\mu$-like track \cite{sandip,enrique}. 

\begin{figure}[t]
\leavevmode
\epsfxsize=2.5in
\epsfysize=2.5in 
\epsfbox{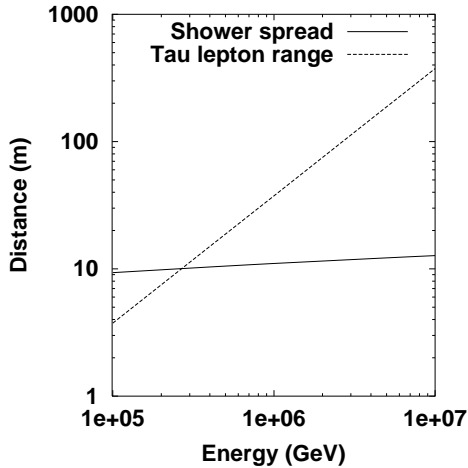}
\caption{Comparison of the tau lepton range and the shower length of the first
         shower in ice/water.}
\end{figure}

The calculation of down ward going contained but separable  
 double shower event rate for a typical km$^{2}$ surface area 
 under water/ice neutrino 
 telescope can be carried out by  replacing 
 the muon range expression with the tauon one and then subtracting  it from the
linear size of a typical high energy neutrino telescope  
 in the event rate formula while using the expected 
$\nu_{\tau}$ flux spectrum given by Eq. (3).
Here, i confine myself by mentioning that the expected number of contained 
but separable double showers
induced by down ward going high energy cosmic tau neutrinos for
 $E\sim 2\cdot 10^{6}$ GeV
may be $ \sim {\cal O}(10) $/yr$\cdot$sr, essentially  irrespective of 
 solution of solar neutrino problem, if one uses the $F^{0}_{e}$ from Ref.
 \cite{szabo} as an example. At this energy,
the two showers initiated by the down ward going high energy cosmic tau 
 neutrinos are well separated (see Fig. 1) such that the amplitude of the 
 second shower is essentially 2 times the first shower.

\section{CONCLUSIONS}

	1. Intrinsically, the flux of high energy cosmic tau neutrinos is quite
small, typically being $F^{0}_{\tau}/F^{0}_{e, \mu}\, <\, 10^{-5}$ from, 
for instance, cosmologically distant cores of Active Galactic Nuclei.

	2. Because of vacuum neutrino flavour oscillations, 
 this ratio can be greatly enhanced. In
the context of three flavour neutrino mixing scheme which can accomodate the 
oscillation solutions to solar and atmospheric neutrino deficits, the final
flux ratio of high energy cosmic neutrinos on earth is 
$F_{e}\sim F_{\mu}\sim F_{\tau} \sim F^{0}_{e}$.

	3. This enhancement in high energy cosmic tau neutrino flux may lead 
to the possibility of its detection in km$^{2}$ surface area under water/ice 
 high energy 
neutrino telescopes. For $2\cdot 10^{6}\leq E$/GeV
$\leq 2\cdot 10^{7}$, the down ward going high energy cosmic tau neutrinos
may produce a double shower signature because of charged current deep 
inelastic scattering followed by a subsequent hadronic decay of associated tau
lepton.

\section*{ACKNOWLEDGEMENTS}

I thank Japan Society for the Promotion of 
Science (JSPS) for financial support.


\begin{thebibliography}{9}
\bibitem{sandip} J. G. Learned and S. Pakvasa, Astropart. Phys. 3 (1995) 267.
\bibitem{taup} See, E. Waxman; V. Berezinsky; P. L. Biermann 
               and G. Sigl in these proceedings.
\bibitem{npb} Athar Husain, Nucl. Phys. B (Proc. Suppl.) 76 (1999) 419
                and references cited therein.

\bibitem{upward} For a recent discussion, see, S. Iyer, M. H. Reno 
                 and I. Sarcevic, hep-ph/9909393 and references cited therein.
\bibitem{moscoso} L. Moscoso, these proceedings.
\bibitem{athar} A some what detailed numerical study supports this order of 
                magnitude estimate; H. Athar, R. A. V\'azquez and E. Zas (in 
                preparation).
\bibitem{szabo} A. P. Szabo and R. J. Protheroe, Astropart. Phys. 2 (1994) 375.
\bibitem{book} Ta-Pei Cheng and Ling-Fong Li (eds.), Gauge theory of 
               elementary particle physics, Claredon Press, Oxford, 1984, 
               p. 411.
\bibitem{pdg} C. Caso et al., The Euro. Phys. J. C3 (1998) 103.
\bibitem{nakahata} M. Nakahata, these proceedings.
\bibitem{enrique} H. Athar, G. Parente and E. Zas (to be submitted). See also
                  scanned transperacies by Athar Husain at URL 
                  http://taup99.in2p3.fr/TAUP99/.
\end{thebibliography}
\end{document}